\documentclass{llncs}
\usepackage[vlined,ruled,linesnumbered]{algorithm2e}
\usepackage{multirow}
\usepackage{floatflt}
\usepackage{amsfonts}
\usepackage{bbold}

\usepackage[pdftex]{graphicx}
\graphicspath{{./fig/}}
\usepackage{amsmath}
\usepackage{subcaption}
\usepackage[caption=false,font=footnotesize]{subfig}
\usepackage{floatflt}
\usepackage{url}
\begin{document}
\pdfoutput=1

%
%
\pagestyle{headings}  
%
\title{Country-scale Exploratory Analysis of Call Detail Records through the Lens of Data Grid Models}
\titlerunning{Country-scale Exploratory Analysis of Call Detail Records}  
%
\author{Romain Guigour\`es\inst{1} \and Dominique Gay\inst{2}  \and\\Marc Boull\'e\inst{2} \and Fabrice Cl\'erot\inst{2} \and Farbice Rossi\inst{3}
}
\authorrunning{R. Guigour\`es et al.} 
\institute{
Zalando\\
Berlin, Germany\\
\and
Orange Labs\\
Lannion, France\\
\and
SAMM EA 4543, Universit\'e Paris 1\\
Paris, France\\
}

\maketitle              

\begin{abstract}
Call Detail Records (CDRs) are data recorded by telecommunications companies, consisting of basic informations related to several dimensions of the calls made through the network: the source, destination, date and time of calls. CDRs data analysis has received much attention in the recent years since it might reveal valuable information about human behavior. It has shown high added value in many application domains like e.g., communities analysis or network planning.\\
In this paper, we suggest a generic methodology for summarizing information contained in CDRs data. The method is based on a parameter-free estimation of the joint distribution of the variables that describe the calls. We also suggest several well-founded criteria that allows one to browse the summary at various granularities and to explore the summary by means of insightful visualizations. The method handles network graph data, temporal sequence data as well as user mobility data stemming from original CDRs data. We show the relevance of our methodology for various case studies on real-world CDRs data from Ivory Coast.
\end{abstract}
%

\section{Introduction}
Telco operators' activities generate massive volume of data, mainly from three sources: networks, service platforms and customers data bases. Particularly, the use of mobile phones generates the so called Call Detail Records (CDRs), containing information about end-point antenna stations, date, time and duration of the calls (the content of the calls is excluded). While this data is initially stored for billing purpose, useful information and knowledge (related to human mobility~\cite{WPS+11,BCH+13}, social interactions and economic activities) might be derived from the large sets of CDRs collected by the operators.

Recent studies have shown the potential added-value of analyzing such data for several application domains: United Nations Global Pulse \cite{GP13} sums up some recent research works on how analysis of CDRs can provide valuable information for humanitarian and development purposes, e.g., for disaster response in Haiti, combating H1N1 flu in Mexico, etc. Also, leveraging country-scale sets of CDRs in Ivory Coast, the recent Orange D4D challenge (Data For Development~\cite{BEC+12}) has given rise to many investigations in several application domains~\cite{VBC+13} such as health improvement, analysis of economic indicators and population statistics, communities understanding, city and transport planning, tourism and events analysis, emergency, alerting and preventing management, mobile network infrastructure monitoring.
Thus, the added-value of analysis of CDRs data does not need to be proved any longer.

Various classical data mining techniques have been applied on CDRs data depending on the features and the task considered: e.g., considering network graphs from (source antenna, destination antenna) data or temporal sequences from (source antenna, date) data appeals for different clustering techniques for summarizing information in the data.

\noindent \textit{Contribution:} in this paper, we suggest an efficient and generic methodology for summarizing CDRs data whatever the features are retained in the analysis. The method is based on data grid models~\cite{Bou11}, a parameter-free joint distribution estimation technique that simultaneously partitions sets of values taken by each variable describing the data (numerical variables are discretized into intervals while the categories of categorical variables are grouped into clusters). The resulting data grid -- that can be seen as a coclustering -- constitutes the summary of the data. The method is thus able to summarize various types of data stemming from CDRs: network graph data, temporal sequence data as well as user mobility data. We also suggest several criteria \textit{(i)} to exploit the resulting data grid at various granularities depending on the needs of analysis and \textit{(ii)} to interpret the results through meaningful visualizations.

\noindent \textit{Outline:} in the next section, we give a brief description of the CDRs data and the various case studies we led on the data. Section~\ref{sec:method} recalls the main principles of data grid models and introduces the criteria for exploiting the resulting data grid. In section~\ref{sec:experiments}, we report the experimental results on the various case studies. Also, we discuss further related work in section~\ref{sec:related} before concluding. 
\section{Data description \& studies}
\label{sec:data}
The CDRs data under study come from the Orange D4D challenge\footnote{\url{http://d4d.orange.com/en/home}} (Data For Development~\cite{BEC+12}). We consider several case studies on two anonymized data sets, namely communication data and mobility data.
%
\subsection{Case studies on communication data}
Communication data consists in 471 millions mobile calls and covers a 5-month period (from 2011, December 1st to 2012, April 28th). The records are described by the four following variables:
\begin{enumerate}
\itemsep0em
\small
\item emitting antenna (1214 categorical values);
\item receiving antenna (1216 categorical values);
\item time of call (with hour precision);
\item date of call (from 2011/12/01 to 2012/04/28);
\end{enumerate}
%

\noindent From this data set, we consider three subsets for:
\begin{enumerate}
\itemsep0em
\item \emph{Analysis of call network between antennas.} Considering emitting antennas, receiving antennas and the calls made between antennas, the data set can be seen as a directed multigraph where nodes are antennas and links are the calls between antennas.

\item \emph{Analysis of output traffic w.r.t. date of call.} We consider emitting antennas and the number of days for each call from referral to first day of recording. This data set can be considered as a temporal event sequence spanning over the whole observation period, where the time is the number of days passed and the events are the emitting antenna IDs.

\item \emph{Analysis of output traffic w.r.t. week day and hour of call.} We consider emitting antennas, the day of the week (stemming from the date and considered as a numerical variable) and the hour of the day for each call. Here the time dimension is represented by two variables and the data of the whole period are folded up to week day and hour.
\end{enumerate}

\subsection{Case studies on mobility data}
Mobility data consists in mobility traces of 50000 users over a 2-week period (from 2012 December 12th to 2012 December 24th), i.e. approximatively 55 millions records. The records are described by the four following variables:
%
\begin{enumerate}
\itemsep0em
\small
\item anonymized user ID (50000 categorical values);
\item connexion antenna (1214 categorical values);
\item time of call (minute precision);
\item date fo call (from 2012/12/12 to 2012/12/24);
\end{enumerate}
%

\noindent From this data, we consider the user trajectories (identified by user ID) inside the network for the following analysis:

\begin{enumerate}
\itemsep0em
%
\item \emph{Analysis of user mobility w.r.t. week day and hour.} We consider the user ID, antennas, week day and hour.
This data set can be considered as a set of spatio-temporal footprints, where each user ID is associated with a sequence of antenna usage over the time dimension. Here again, the time dimension is represented by two variables and the data of the whole period is folded up to week day and hour.
\end{enumerate}
\section{Exploitation of data grid models}
\label{sec:method}
Data grid models aim at estimating the joint distribution between several variables of mixed-types (categorical as well as numerical). The main principle is to simultaneously partition the values taken by the variables, into groups/clusters of categories for categorical variables and into intervals for numerical variables. The result is a multidimensional ($K$-D) data grid whose cells are defined by a part of each partitioned variable value set. Notice that in all rigor, we are working only with partitions of variable value sets. However, to simplify the discussion we will sometime use a slightly incorrect formulation by mentioning a ``partition of a variable'' and a ``partitioned variable''.

In order to choose the ``best'' data grid model $M^{\ast}$ (given the data) from the model space $\mathcal{M}$, we use a Bayesian Maximum A Posteriori (MAP) approach. We explore the model space while minimizing a Bayesian criterion, called cost. The cost criterion implements a trade-off between the accuracy and the robustness of the model and is defined as follows:
%
\begin{align}
cost(M) = -\log(\underbrace{p(M \mid D)}_{\textrm{posterior}}) \propto  -\log(\underbrace{p(M)}_{\textrm{prior}} \times \underbrace{p(D \mid M)}_{\textrm{likelihood}})\nonumber
\end{align}
%
Thus, the optimal grid $M^{\ast}$ is the most probable one (maximum a posteriori) given the data. Due to space limitation, the details about the $cost$ criterion and the optimization algorithm (called \textsc{khc}) are reported in appendix. Hereafter, we focus on the tools for exploiting the grid and their applications on large-scale CDRs data. The key features to keep in mind are: \textit{(i)} \textsc{khc} is parameter-free, i.e., there is no need for setting the number of clusters/intervals per dimension; \textit{(ii)} \textsc{khc} provides an effective locally-optimal solution to the data grid model construction efficiently, in sub-quadratic time complexity ($O(N\sqrt{N}\log N)$ where $N$ is the number of data points).

\subsection{Data grid exploitation and visualization}
\label{subsec:gridexploitation}
Because of the very large number observations in CDRs data, the optimal grid $M^{\ast}$ computed by \textsc{khc} can be made of hundreds of parts per dimension, i.e., millions of cells, which is difficult to exploit and interpret. To alleviate this issue, we suggest a grid simplification method together with several criteria that allow us to choose the granularity of the grid for further analysis, to rank values in clusters and to gain insights in the data through meaningful visualizations.\\

\noindent \textbf{Dissimilarity index and grid structure simplification.}
We suggest a simplification method of the grid structure that iteratively merge clusters or adjacent intervals -- choosing the merge generating the least degradation of the grid quality. To this end, we introduce a dissimilarity index between clusters or intervals which characterize the impact of the merge on the $cost$ criterion.

\begin{definition}[Dissimilarity index]
\label{def:diss}
Let $c_{.1}$ and $c_{.2}$ be two parts of a variable partition of a grid model $M$. Let $M_{c_{.1}\cup c_{.2}}$ be the grid after merging $c_{.1}$ and $c_{.2}$. The dissimilarity $\Delta(c_{.1},c_{.2})$ between the two parts $c_{.1}$ and $c_{.2}$ is defined as the difference of $cost$ before and after the merge:
%
\begin{align}
\Delta(c_{.1},c_{.2}) = cost(M_{c_{.1}\cup c_{.2}}) - cost(M)
\end{align}
\end{definition}
%
%
When merging clusters that minimize $\Delta$, we obtain the sub-optimal grid $M'$ (with a coarser grain, i.e. simplified) with minimal $cost$ degradation, thus with minimal information loss w.r.t. the grid $M$ before merging. Performing the best merges w.r.t. $\Delta$ iteratively over the $K$ variables without distinction, starting from $M^{\ast}$ until the null model $M_{\emptyset}$, $K$ agglomerative hierarchies are built and the end-user can stop at the chosen granularity that is necessary for the analysis while controlling either the number of clusters/cells or the information ratio kept in the model. The information ratio of the grid $M'$ is defined as follows:
\vspace{-0.2cm}
\begin{align}
\label{eq:informationratio}
IR(M') = \frac{cost(M')-cost(\mathcal{M_{\emptyset}})}{cost(M^*) - cost(M_{\emptyset})}
\end{align}
where $M_{\emptyset}$ is the null model (the grid with a single cell).\\

%

\noindent \textbf{Typicality for ranking categorical values in a cluster.} When the grid is coarsen during the hierarchical agglomerative process, the number of clusters per categorical dimension decreases and the number of values per cluster increases. It could be useful to focus on the most representative values among thousands of values of a cluster. In order to rank values in a cluster, we define the typicality of a value as follows.

\begin{definition}[Typical values in a cluster]
\label{def:typicality}
For a value $v$ in a cluster $c$ of the partition $Y^M$ of dimension $Y$ given the grid model $M$, the typicality of $v$ is defined as:
\begin{equation}
\label{eq:typicality}
\begin{array}{l}
\tau (v,c) = \\
\frac{1}{1-P_{Y^M}(c)} \times  \nonumber \\
\sum_{c_j\in Y^M\atop c_j\neq c} P_{Y^M}(c_j)(cost(M|c\setminus v,c_j\cup v)-cost(M)) \nonumber
\end{array}
\end{equation}
where $P_{Y^M}(c)$ is the probability of having a point with a value in cluster $c$, $c\setminus v$ is the cluster $c$ from which we have removed value $v$, $c_j\cup v$ is the cluster $c_j$ to which we add value $v$ and $M|c\setminus v,c_j\cup v$ the grid model $M$ after the aforementioned modifications.
\end{definition}

Intuitively, the typicality evaluates the average impact in terms of $cost$ on the grid model quality of removing a value $v$ from its cluster $c$ and reassigning it to another cluster $c_j\neq c$. Thus, a value $v$ is representative (say typical) of a cluster $c$ if $v$ is ``close'' to $c$ and ``different in average'' from other clusters $c_j\neq c$. Notice that this measure does not introduce any numerical encoding of the categories of the categorical variable under study.\\

%
%

\noindent \textbf{Insightful visualizations with Mutual Information.}
It is common to visualize 2D coclustering results using 2D frequency matrix or heat map. For $K$D coclustering, it is useful to visualize the frequency matrix of two variables while selecting a part of interest for each of $K-2$ other variables. We also suggest an insightful measure for co-clusters to be visualized, namely, the Contribution to Mutual Information (CMI) -- providing additional valuable visual information inaccessible with only frequency representation. Notice that such visualizations are also valid whatever the variable of interest.

\begin{definition}[Contribution to mutual information]
\label{def:cmi}
Given the $K-2$ selected parts $c_{i_3\ldots i_K}$, the mutual information between two partitioned variables $Y_1^{M}$ and $Y_2^{M}$ (from the partition $M$ of $Y_1$ and $Y_2$ variables induced by the grid model $M$) is defined as:
\begin{align}
\label{eq:cmi}
MI(Y_1^{M};Y_2^{M})=\sum_{i_1=1}^{J_1}\sum_{i_2=1}^{J_2}MI_{i_1i_2}\nonumber \\
\text{where} \quad MI_{i_1i_2}=p(c_{i_1i_2})\log \frac{p(c_{i_1i_2})}{p(c_{i_1})p(c_{i_2})}
\end{align}
where $MI_{i_1i_2}$ represent the contribution of cell $c_{i_1i_2}$ to the mutual information.
\end{definition}

Thus, if $MI_{i_1i_2} > 0$ then $p(c_{i_1i_2}) > p(c_{i_1.})p(c_{.i_2})$ and we observe an excess interaction between $c_{i_1.}$ and $c_{.i_2}$ located in cell $c_{i_1i_2}$ defined by parts $i_1$ of $Y_{1}^M$ and $i_2$ of $Y_{2}^M$. Conversely, if $MI_{i_1i_2} < 0$, then $p(c_{i_1i_2}) < p(c_{i_1.})p(c_{.i_2})$, and we observe a deficit of interactions in cell $c_{i_1i_2}$. Finally, if $MI_{i_1i_2} = 0$, then either $p(c_{i_1i_2})=0$ in which case the contribution to MI and there is no interaction or $p(c_{i_1i_2}) = p(c_{i_1.})p(c_{.i_2})$ and the quantity of interactions in $c_{i_1i_2}$ is that expected in case of independence between the partitioned variables.

The visualization of cells' CMI highlight valuable information that is local to the $K-2$ selected parts and bring complementary insights to exploit the summary provided by the grid. In our experiments, we show the added-value of those visualizations on CDRs data from Ivory Coast described in Section~\ref{sec:data}.

\section{Exploration results}
\label{sec:experiments}

This section describes the application of the previously introduced exploratory analysis framework on mobile data. 
Each application of \textsc{khc}\footnote{available at \url{http://www.khiops.com}} on  the case study data is achieved within a day of computation -- which confirms the efficiency of the method. First, we apply the co-clustering with two categorical variables to build clusters of antennas based on the mobile traffic. Then, we study the time evolution of the calls distribution using data grid models with one categorical and one continuous variables. Next, we extend the previous analysis by applying our triclustering technique on the antennas, the weekday and the daytime in order to track active and inactive areas in function of the day and the hour. Finally, we investigate on the users behavior according to the antennas they use, the weekday and the time. This last study is an application of data grid models in four dimensions, i.e a tetra-clustering.
\subsection{The clusters}
\label{subsec:clusters}
The application of data grid models on the call detail records provides a segmentation with $1150$ clusters, that corresponds to nearly one antenna per cluster. This is due to the large amount of data -- 471 millions CDRs. Indeed the number of calls is so high for each antenna that the distribution of calls originating from (resp. terminating to) each antenna can be distinguished from each other. 

\begin{floatingfigure}[r]{6.5cm}
\centering
\includegraphics[width=.5\columnwidth]{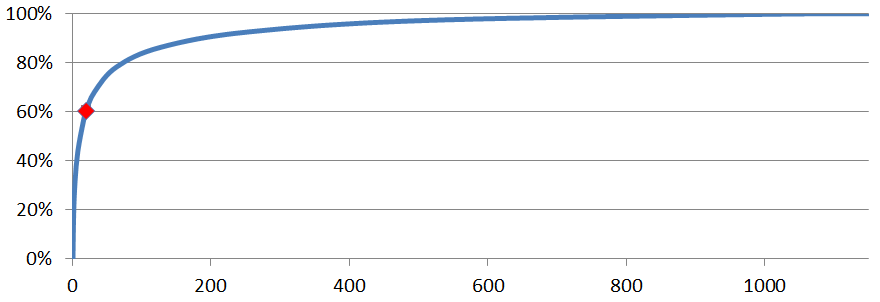}
\vspace{-0.2cm}
\caption{\scriptsize Evolution of the information kept in the data grid model w.r.t. the number of clusters using the ascending hierarchical post-processing -- from optimal data grid $M^{\ast}$ (100\%) to the null model $M_{\emptyset}$ (0\%).}
\label{fig:pareto}
\end{floatingfigure}
In order to obtain a more interpretable segmentation, we apply the post-treatment introduced in the Section~\ref{subsec:gridexploitation}. Figure~\ref{fig:pareto} plots the information ratio (see definition~\ref{def:diss}) versus the number of clusters for all intermediate models obtained during the ascending hierarchical post-processing. Interestingly, the resulting Pareto curve shows that very informative models are obtained with few clusters. In our study, we decrease the number of clusters until keeping $60\%$ of the model informativity. By doing this, we obtain $20$ clusters that is a satisfying number for the interpretation.


Throughout the simplification process, both partitions of source and target antennas stay identical. Thus we consider only the partition of source antennas for the rest of the study. We have plotted the clusters on a map in Figure~\ref{fig:ci_map}. Antennas are identified using dots, which color match with the cluster they belong to.\\
The first observation is the strong correlation between the clusters and the geography of the country. Indeed, antennas from a same cluster are close to each other. The size of the clusters is almost the same in terms of area and match with the administrative zones of the country. There is an exception for Abidjan that is split into four clusters. This is due to the high concentration of antennas in the city ($32\%$ of the ivorian antennas) and the dense phone traffic ($34\%$ of the calls).

We use the typicality (see definition~\ref{def:typicality}) to rank the antennas of each cluster. The place, where the antenna with the highest typicality is located, is used to label the cluster. On the map in Figure~\ref{fig:ci_map}, the size of the dots are proportional to the antenna typicality. Most typical antennas are located in the main cities of Ivory Coast. This phenomenon has already been observed in \cite{Krings2010} and \cite{Guigoures2011}: the clusters match with the area of influence of the main cities of a country. We note some exceptions. Among them, the cluster of the city of Sassandra contains the antennas of the city of Divo, while Divo is almost $4$ times bigger than Sassandra (population wise) and is the sixth Ivorian city. Antennas in Divo are $40\%$ less typical than the ones in Sassandra, meaning that allocating them to another cluster would be less costly for the criterion. Actually, calls emitted from Divo are significant in direction to other regions of Ivory Coast whereas calls from Sassandra are more internal to its region. In more formal terms, the calls distributions of the antennas in Divo are closer to the marginal distribution than to its cluster's distribution. This observation is not really surprising because Divo has experienced a recent growth of its population, due to migrations within the country \cite{Gnabeli2008}. Divo is also located in an area specialized in the intensive farming, that attracts seasonal workers from other parts of Ivory Coast.



\begin{floatingfigure}[r]{6.5cm}
\centering
\includegraphics[width=.5\columnwidth]{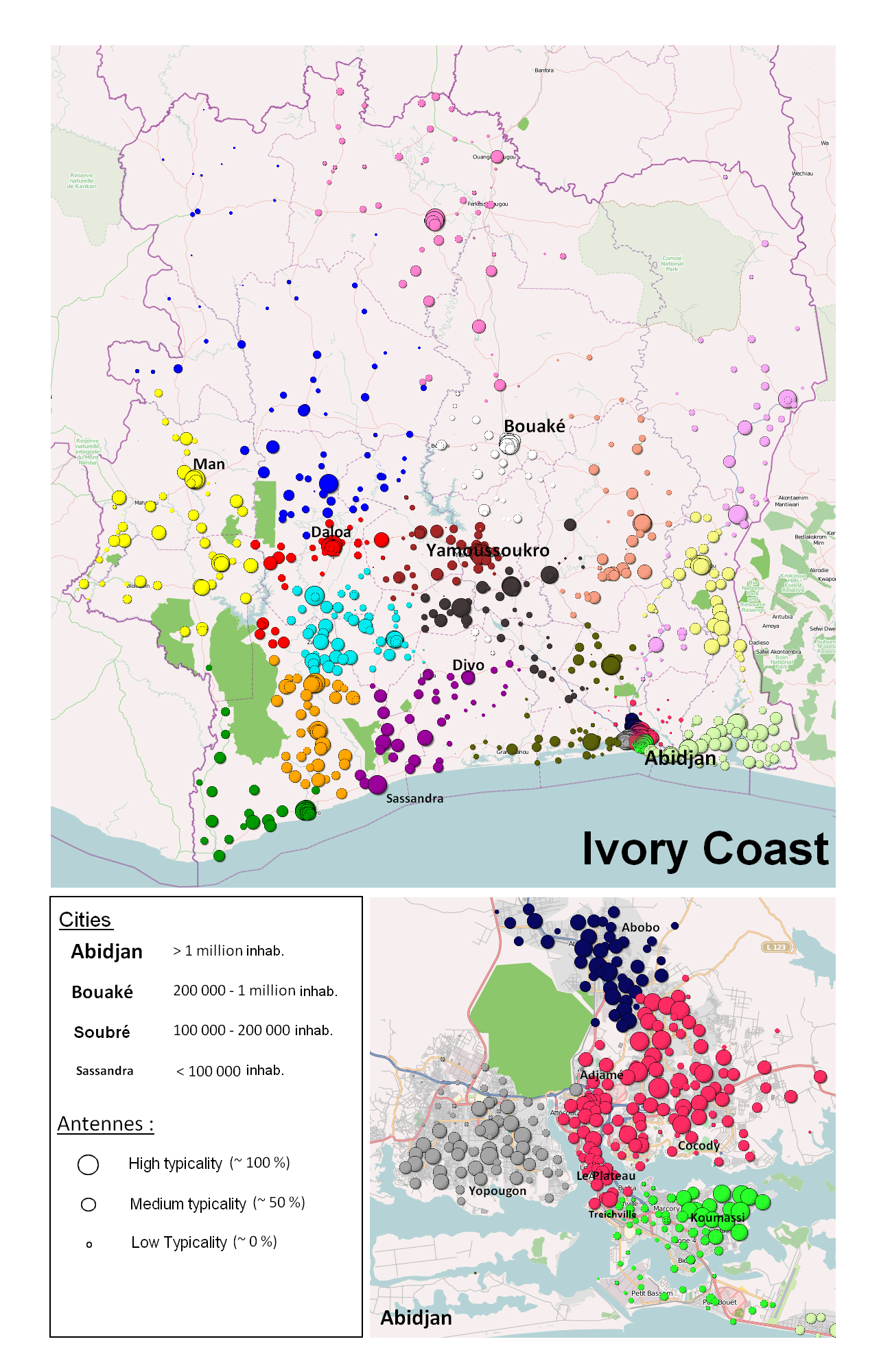}
\caption{\scriptsize Twenty clusters displayed on Ivory Coast map. There is one color per cluster.}
\label{fig:ci_map}
\end{floatingfigure}
Finally, let us focus on the segmentation of Abidjan. The city is divided into four parts with a strong socioeconomic correlation. The first cluster -- in red in Figure~\ref{fig:ci_map} -- covers central Abidjan, including the Central Business District (le Plateau), the transport hub (Adjam\'e) and the embassies and upper class area (Cocody). The second cluster -- in light green in the Figure~\ref{fig:ci_map} -- is located in the South of the city. The covered neighborhoods are mainly residential areas and ports. Note that this cluster and the previous one are separated by a strip of sea, except for its North part that is included in the previous cluster. This very localized neighborhood matches with the party area of Abidjan.
Finally, the last two clusters group antennas located in two areas with a similar profile: these are lower class neighborhoods. These clusters are separated not only because they are located in different parts of the city but especially because their call distribution differs: Abobo in dark blue and Yopougon in grey in the Figure~\ref{fig:ci_map}.

\subsection{Traffic between clusters}
\label{subsec:traffic_clusters}

In the Section~\ref{sec:method}, we have introduced the contribution to the mutual information. We propose to use it in order to visualize the lacks and excesses of calls between the clusters, compared to the expected traffic in case of independence. Whatever the granularity level of the clustering, we observe a strong excess of calls from the clusters to themselves and way weaker excesses and lacks between clusters. Studying the traffic within the clusters has a limited interest. We only focus on the inter-clusters traffic. To visualize the traffic between clusters, we use a finer clustering than in Section~\ref{subsec:clusters}. Here we have 355 clusters for $95\%$ informativity (see Figure~\ref{fig:pareto}).

\begin{figure}[t]
\centering
\includegraphics[width=.5\columnwidth]{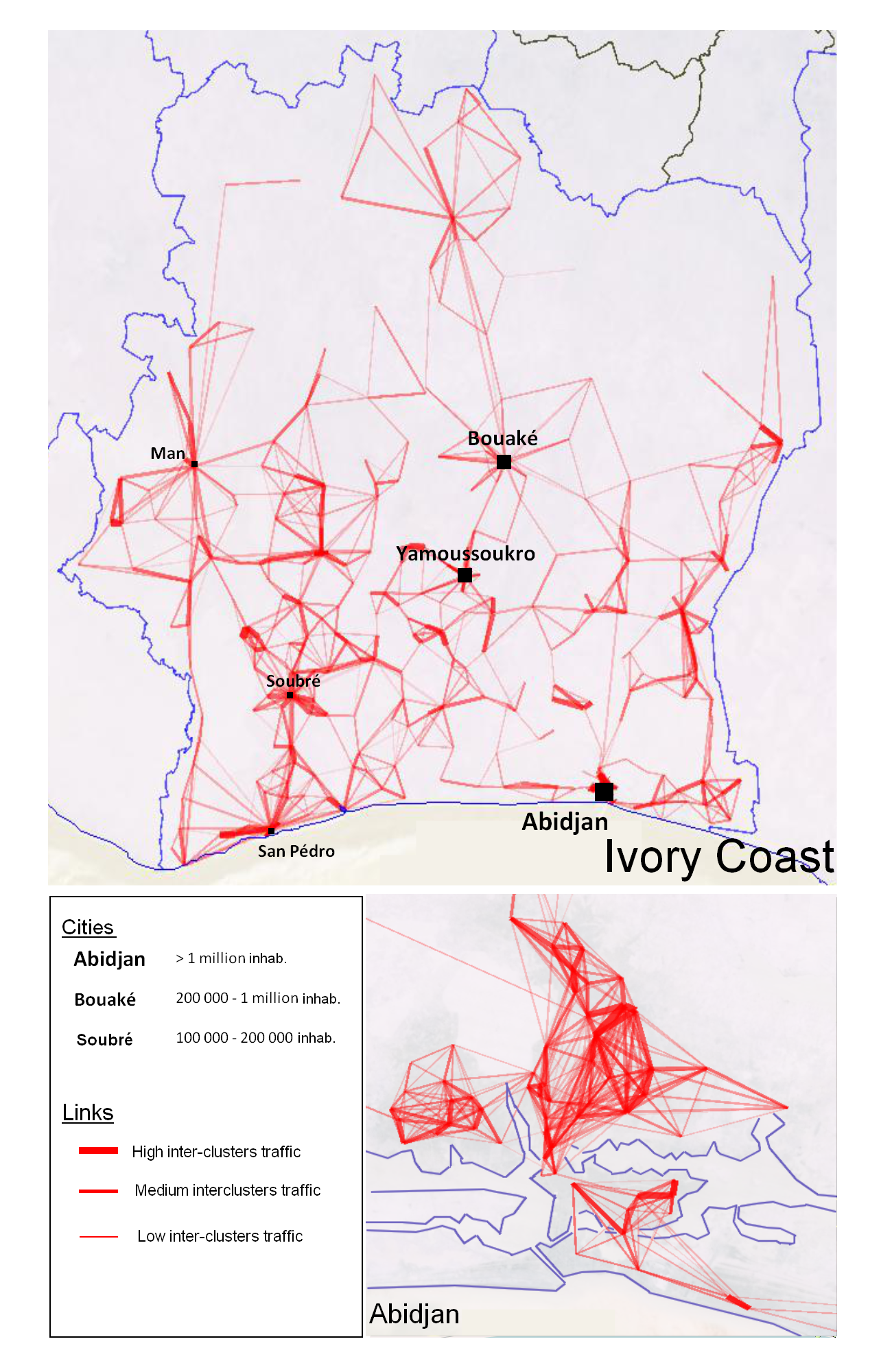}
\vspace{-0.2cm}
\caption{\scriptsize Excess of calls between clusters of antennas}
\label{fig:mi_map}
\end{figure}

In Figure~\ref{fig:mi_map}, the red segments on the map are the excesses of traffic between clusters. The end points of the segments are drawn at the positions of the most representative antennas of the associated clusters (i.e with the highest typicalities). The opacity of a segment is proportional to the value of the contribution to mutual information and its width is proportional to the number of calls between clusters. The biggest cities -- like Bouak\'e, San Pedro and Man -- are clearly marked on the map: they are regional capitals, a fact that is highlighted by the call traffic visualization. The case of Bouak\'e is particularly interesting. Although it is not the country capital, its national influence seems bigger than the one of Yamoussoukro, the actual capital. Yamoussoukro is twice smaller than Bouak\'e (population wise) and is a quite recent city where there is no major economical activity, contrary to Bouak\'e. This fact can explain our observation.

Excess of traffic between major cities is a rare phenomenon. Cities are more like phone hubs, except in the West of the country around Soubr\'e. This area is not a densely populated area but corresponds to a region with important migration flows. Finally, in Abidjan we can note important excesses of traffic within neighborhoods, but not between neighborhoods.

\subsection{Temporal Analysis of the Calls Distribution}
\label{subsec:temp_cd}
In this analysis, we track the evolution of the traffic over time. The studied time period runs from 2011, December 1st to 2012, April 28th. The model that is introduced in Section~\ref{sec:method} has been designed to deal with several variables, either continuous or categorical.
Thus we could study the calls from emitting antennas to destination antennas according to the time, using three variables. However in the Section~\ref{subsec:clusters}, we have shown that the correlation between source and destination antennas is very high. The evolution of the calls distribution over time might be the same for both sets of antennas. Therefore, we only study the time evolution of the originating calls: one call is described by the emitting antenna and a day count (stemming from the date).


\begin{floatingfigure}[r]{6cm}
\centering
\includegraphics[width=.5\columnwidth]{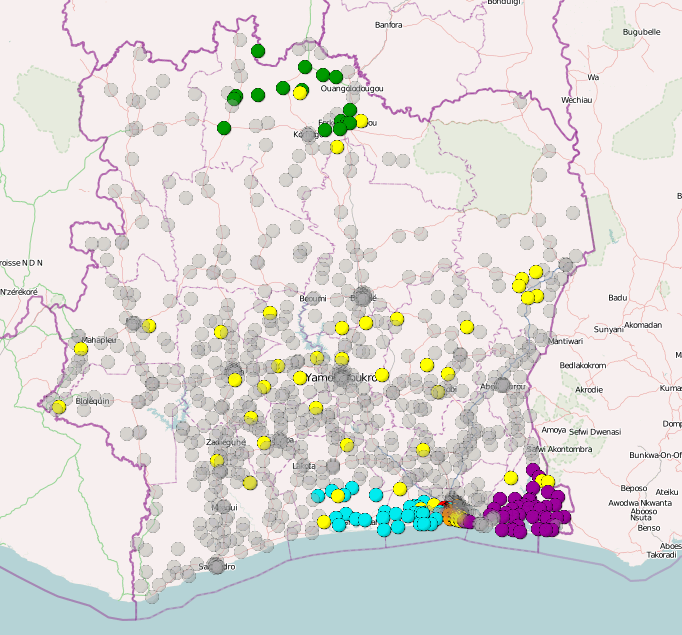}
\caption{\scriptsize Antennas activity clusters projected on Ivory Coast map. Colored clusters show inactivity periods while grey clusters indicate antennas whose traffic is complete over the period.}
\label{fig:activity_map}
\end{floatingfigure}
In this study, the clustering of antennas is also too fine for an easy interpretation (1051 clusters of antennas and 140 intervals for the day count). We apply the same post-treatment than in the Section~\ref{sec:method}, so that the informativity of the model is $80\%$, with ten clusters of antenna and twenty time segments. The main problem of this analysis is the missing data. Indeed, some antennas emitted no call during some time periods. Consequently, we obtain time segments that are strongly correlated with missing data. For the same reason, antennas are grouped because they experienced an absence of calls during one or several same periods.

In the Figure~\ref{fig:activity_map}, the colored antennas belong to clusters having experienced simultaneous absences of calls. We note that the green, orange, light blue and purple clusters are located in localized area. The missing data are during short periods for these clusters. This grouping might be due to localized technical issues on the network. The antennas of the yellow cluster are spread over the country. These antennas are grouped because they have been activated at the same date. This use case provides a better understanding the dysfunctions in the network over the year.
\subsection{Output communications w.r.t. week day and hour}
\label{subsec:output_dh}
In this analysis with use three variables to describe the calls: the emitting antenna, the week day and hour. Our objective is to build simultaneously a partition of the antennas, a partition of the week days and a discretization of the hour. This approach is a triclustering and for the same reasons as previously, we only keep the emitting antennas.

At the finest level, we obtain a triclustering with $806$ clusters of emitting antennas, $7$ clusters of days and $22$ time segments. These results must be simplified to ease the interpretation. Here we fix the numbers of clusters of days and time segments, since they are acceptable for the analysis. We only reduce the number of clusters of antennas. With four clusters of antennas, we keep $51\%$ of the informativity of the model. 

\begin{floatingfigure}[r]{6.5cm}
\centering
\includegraphics[width=.5\columnwidth]{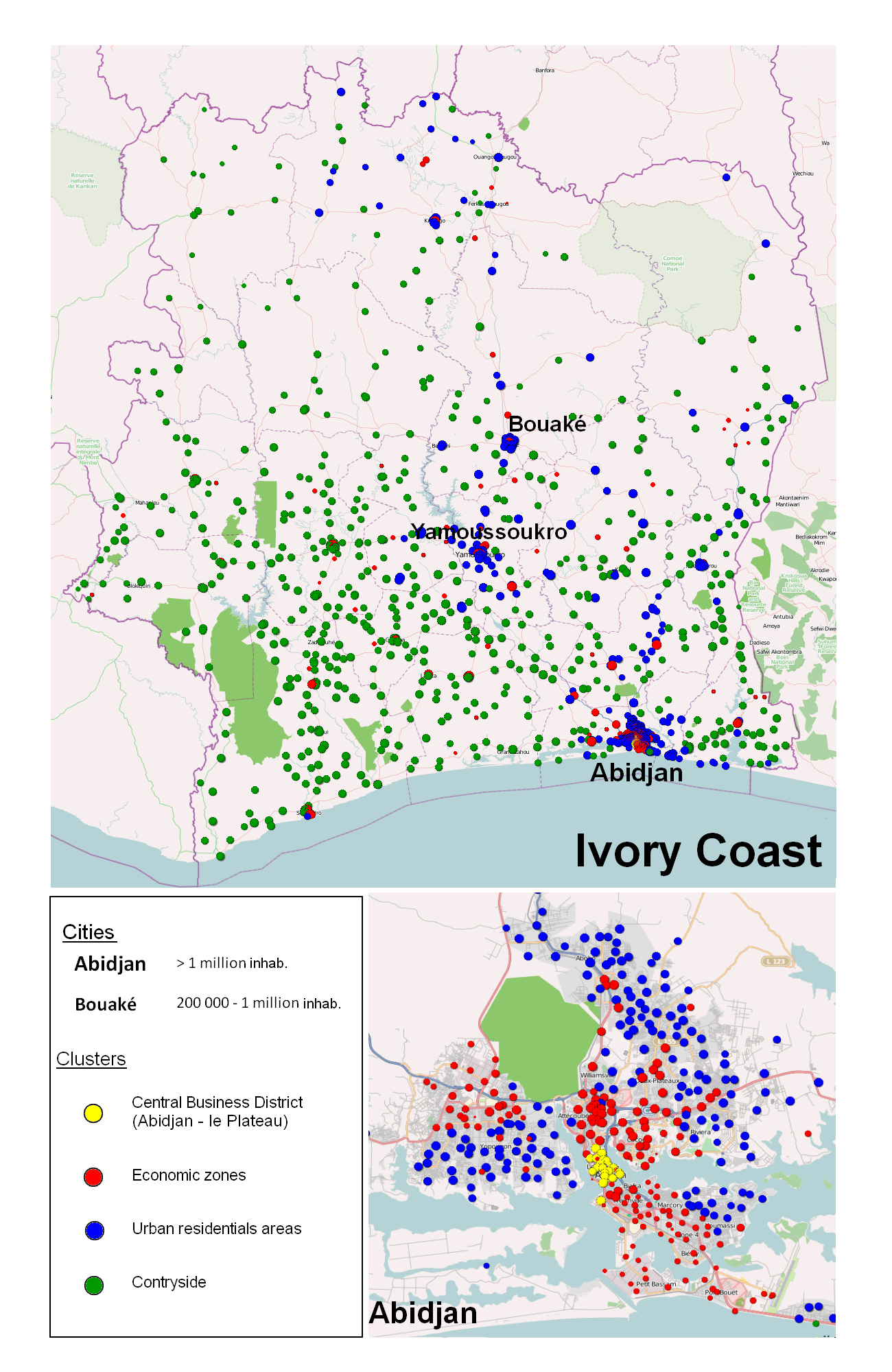}
\caption{\scriptsize Clusters on the map of Ivory Coast. Dots are antennas. There is one color per cluster.}
\label{fig:carteWeekHour}
\end{floatingfigure}
Antennas are displayed on the map of Figure~\ref{fig:carteWeekHour}. We also build a calendar (see Figure~\ref{fig:calWeekHour}) for each cluster with days in columns and time segments in lines. The color of the cells indicates the excesses (red) or the lacks (blue) of traffic emitted from the corresponding cluster. The lacks and excesses are measured using the contribution to the mutual information (see definition~\ref{def:cmi}) between the cluster and the cross product of the cluster of weekday and the time segment: $MI(X_1^M;X_2^M \times X_3^M)$, with $X_1^M$ the partition of the antennas, $X_2^M$ the partitions of the weekdays and $X_3^M$ the discretization of the time.

Let us make an analysis of each cluster of antennas:\\

\noindent{\textbf{Abidjan - Le Plateau (yellow cluster)}
This cluster covers exactly the Central Business District of Abidjan. In the calendar of Figure~\ref{fig:calWeekHour}, we observe an excess of calls from the Monday to the Friday, between 8-9am and 4-5pm. The rest of the time, there is a low lack of traffic emitted from this area. This means that during the office hours the phone traffic is higher than expected and lower the rest of the time. This is representative of this type of area: a non-residential business district.\\

\noindent{\textbf{Economic zones (red cluster)}
The antennas of this cluster are located either in the commercial areas of the cities or in areas with a strong economic activity, like plantations or mines. In Abidjan, these antennas are located in industrial zones (South and North-West), the shopping districts (North of the business district) and the universities and embassies neighborhood (East). The traffic in these areas is mainly in excess from the Monday to the Saturday between 9 am and 5 pm. The correlation is very strong between the working hours and the calls traffic on these areas.\\

\noindent{\textbf{Urban residential areas (blue cluster)}}
The antennas belonging to this cluster are mainly located in the cities like Abidjan, Bouak\'e and Yamoussoukro. If we focus on Abidjan, we realize that the cluster covers the residential neighborhood located in the West and in the North-East of the city. At a finer level of partition of the antennas, this cluster would be split according to the socioeconomic class of the neighborhood: the upper class neighborhood in the East of the city is separated from the lower class neighborhoods, located in the North and the West. The calendar shows lacks of calls during the office hours and excesses the weekend, the night and the early morning during the week. This is correlated with the presence of people in residential areas. Note that the excesses of calls start around 8 pm, while it stops around 5 pm in the Central Business district or in economic areas. This time lag is due to the cheaper price of calls after 8 pm.\\

\noindent{\textbf{The countryside (green cluster)}}
The antennas of this cluster are spread over the country, except in Abidjan and other cities in general. The calendar for this cluster is quite similar to the one of the urban residential areas, except that the excess periods are limited to the early evening and the whole Sunday. 

\begin{figure}[t]
  \centering
  \begin{subfigure}[h]{.45\linewidth}
			\includegraphics[width=\columnwidth]{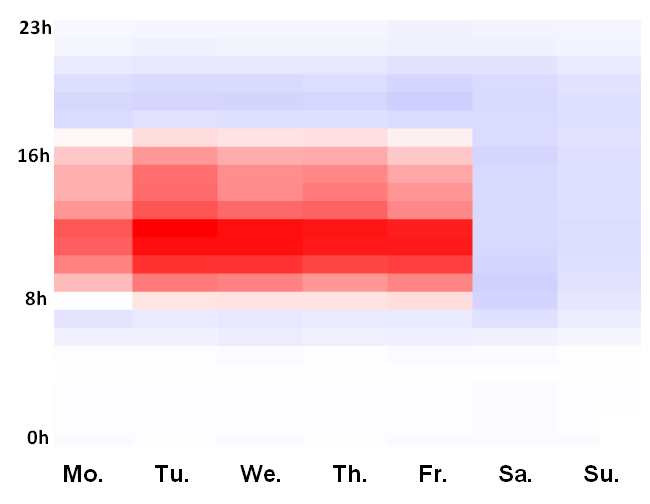}
			\vspace{-0.2cm}
			\caption{\scriptsize Le Plateau (Abidjan)}
	\end{subfigure}
	\begin{subfigure}[h]{.45\linewidth}
			\includegraphics[width=\columnwidth]{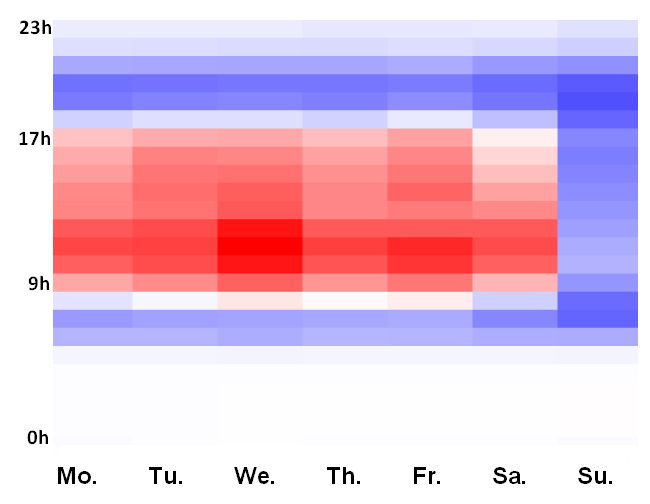}
			\vspace{-0.2cm}
			\caption{\scriptsize Activity areas}
	\end{subfigure}
	\\
	\begin{subfigure}[h]{.45\linewidth}
			\includegraphics[width=\columnwidth]{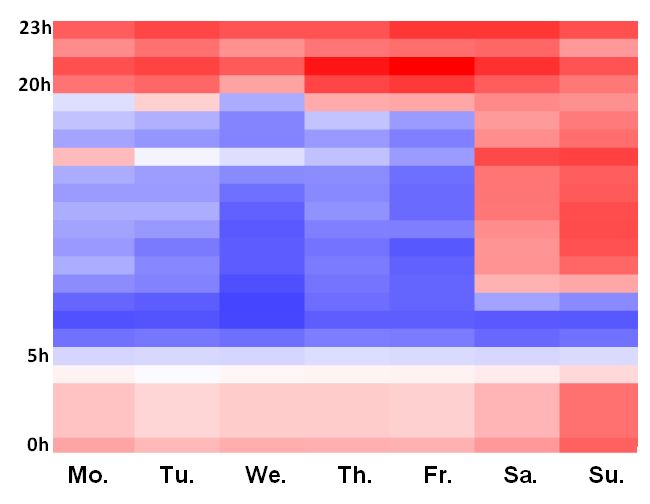}
			\vspace{-0.2cm}
			\caption{\scriptsize Urban residential areas}
	\end{subfigure}
	\begin{subfigure}[h]{.45\linewidth}
			\includegraphics[width=\columnwidth]{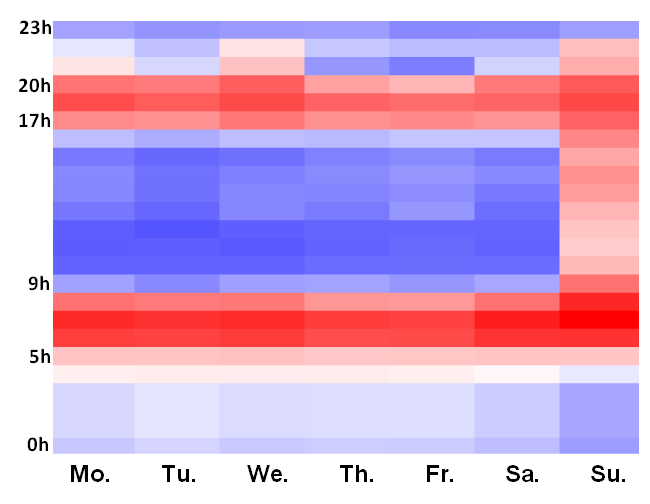}
			\vspace{-0.2cm}
			\caption{\scriptsize Countryside}
	\end{subfigure}
\vspace{-0.2cm}
\caption{\scriptsize Calendars of excesses (red) and lacks of calls emitted from each of the four clusters of antennas, in function of the weekday and the daytime.}
  \label{fig:calWeekHour}
\end{figure}

\subsection{User mobility analysis w.r.t. week day and hour}
\label{subsec:mobility_dh}
The data of this study are customers trajectories. For a set of $50000$ anonymized users, we have the antennas and the timestamps with their uses of the network. The data are the connections to the network, that are identified by a user identifier, an antenna, a week day and hour. In this study, we apply a tetra-clustering in order to build simultaneously clusters of users, of antennas, of week days and a discretization of the daytime. Here two users have the same profile if they connected to the same groups of antennas, the same days of the weeks and at the same time periods. The data are filtered so that only the most mobile users are kept. A mobile user is characterized by a frequent use of a large set of distinct antennas. After filtering, we keep $6894$ users.

At the finest level, we have $237$ clusters of users, $218$ clusters of antennas and three time segments. Week days are not grouped: each of them is in its own cluster. This clustering is too fine for an easy interpretation, so we use the post-treatment, introduced in Section~\ref{sec:method} to simplify the model. We keep $50\%$ of informativity, that enables a reduction of the numbers of clusters of users and antennas to $40$, and the numbers of groups of week days and hour segments to two. The week is divided in two parts: the working days and the weekend. As for the hour, the split occurs around 6 pm. The intervals are 0 am - 6 pm and 6 pm - 12 am. Note that the bound at midnight is artificial, because the day start as this time. The cut at 6 pm is the last in the hierarchy of the time segmentation. Then it would have been more relevant to consider a day from 6 pm to 6 pm the next day. Nevertheless, it is easier to have an interpretations on a ``usual'' time period between 0 am and 12 pm. Therefore we keep the following segmentation: 0 am - 6 pm, 6 pm - 12 pm.

\begin{figure}[htbp!]
  \centering
  \begin{subfigure}[b]{.45\columnwidth}
			\includegraphics[width=\columnwidth]{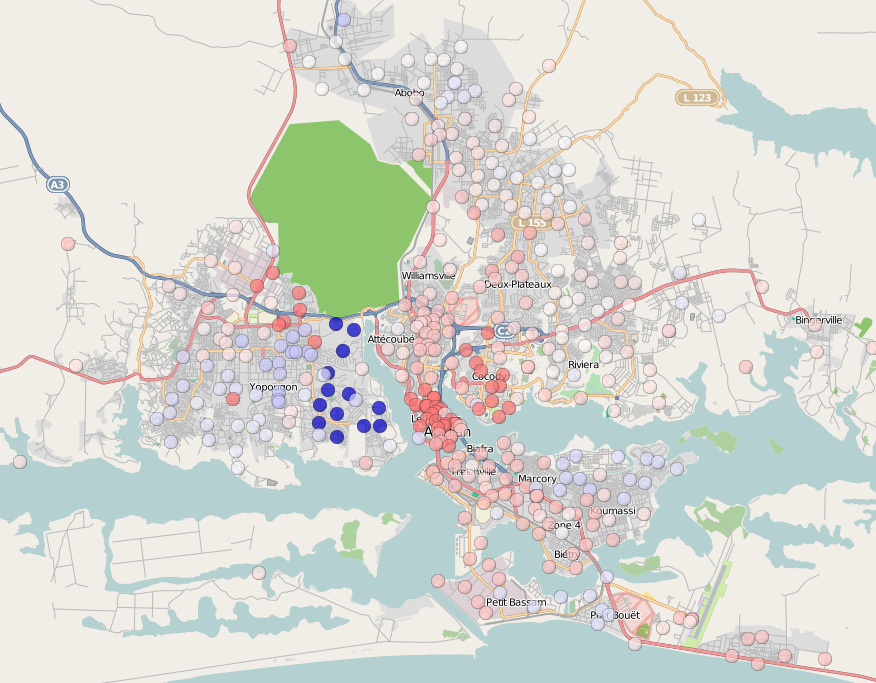}
			\caption{\scriptsize Working days before 6 pm}
	\end{subfigure}
	\hfill
	\begin{subfigure}[b]{.42\columnwidth}
			\includegraphics[width=\columnwidth]{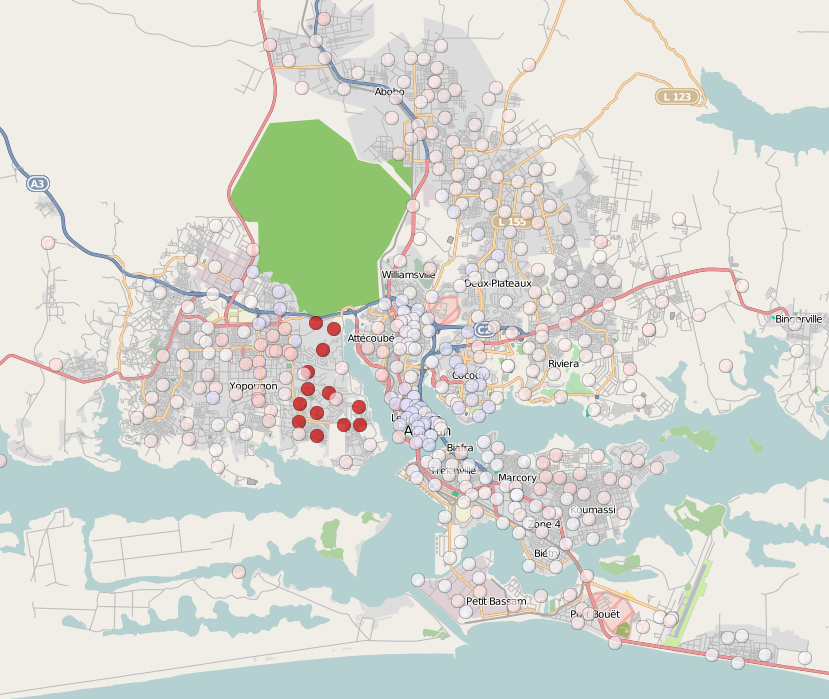}
			\caption{\scriptsize Working days after 6 pm}
	\end{subfigure}
	\\
\caption{\scriptsize For a group of user, excesses and lacks of uses of antennas according to the day of the week and the time of the day. Focus on Abidjan.}
  \label{fig:CItetra}
\end{figure}

We aim at characterizing the users behaviors in terms of mobility. We focus on a group of users to illustrate our results. The maps of Figure~\ref{fig:CItetra} are the maps of excesses and lacks of traffic in Abidjan during the week, for both periods of the day and for a selected group of users. The colors correspond to the mutual information $MI(X_1^M ; X_2^M \times X_3^M \times X_4^M)$ where $X_1^M$ is the partition of antennas; $X_2^M$, the partition of the weekdays; $X_3^M$ the discretization of the daytime; and $X_4^M$, the partition of the users.

Users of the studied cluster mainly connect to the antennas located in the East of Abidjan after 6 pm during the working days, while they rarely connect to the same antennas before 6 pm the same days. Then it can be assumed that the selected cluster of users groups people living in the same area. This hypothesis is reinforced by the socioeconomic nature of this part of Abidjan: it is a residential area. The contributions to mutual information of the other clusters of antennas are smaller. Three areas experience excesses of traffic before 6 pm and lacks after 6 pm. They correspond to the business district (Le Plateau), the embassies and universities neighborhood and the industrial zone located in the West of the city. The common point of all these areas is their economic activity during the day. To sum up, we can assume that the users of the selected cluster are similar in that they live in the same area and work during the week in three localized area of Abidjan.

\subsection{Synthesis}
This section aimed at illustrating how a co-clustering approach could be useful to extract different information on a single telephone data set. We have also shown that some mathematically defined exploratory analysis concepts helped us to make an interpretation of our results. In the first part of the analysis, we have shown how people from the same zone tend to call the same areas of the country. Using mutual information between clusters of antennas, we are able to plot the network of calls in the country. Finally, we made a temporal analysis of the emitted calls, highlighting the differences of behaviors of mobile users according to the area where they live in the country. In a last study of anonymized user mobility data, we proposed a tetra-clustering (or co-clustering in four dimensions) and focused on one cluster of users to show how we can build users profile and investigate their mobile usage. These results have been discussed and the interpretations have been validated by a sociologist from the University of Bouak\'e in Ivory Coast.\\
\noindent \textbf{Impacts on economic strategy.} Besides the high-level knowledge extracted from country-scale data and confirmed by local sociologists, these studies have also a strong impact on future economic development strategy, mainly in two identified branches:
\vspace{-0.1cm}
\begin{itemize}
\itemsep0em
\item \emph{Network planning strategy:} In 2014, there are around 20M inhabitants in Ivory Coast and the mobile service penetration rate is $\simeq 84\%$ -- with a still growing mobile phone market. Maps resulting from the first case study (that can be seen as the network of calls available at various granularities, see Section~\ref{subsec:clusters}) are considered as an additional input for network planning and investment; for instance to help network designer in answering questions about how many and where the next antennas have to be set while preserving the quality of service at a reasonable cost.

\item \emph{Yield management pricing strategy:} a part of the pricing policy, called \textit{Bonus Zone}, established in Ivory Coast offers discount prices (from 10\% to 90\%) to calling users depending on the location and hour of the emitting call. Maps and calendars resulting from case studies (Sections~\ref{subsec:output_dh} and~\ref{subsec:mobility_dh}) that are available at various granularities, provide valuable information to economic analysts in order to design optimized spatio-temporal pricing policy in \textit{Bonus Zone} context.
\end{itemize}
\vspace{-0.2cm}


\section{Related Work}
\label{sec:related}
CDRs data have received much attention in recent years. Famous applications of CDRs data analysis are for the benefit of social good: e.g., in the transportation domain, \cite{BCL+13} suggest a system for public transport optimization.
%
Mobile phones may also provide other types of data (e.g., the Nokia Mobile Data Challenge~\cite{LGA+13}), like applications events, WLAN connection data, etc. For instance, \cite{JFY+13} pre-processed phone activities of one million users to obtain information about their approximative temporal location, then mined daily motifs from the spatio-temporal data to infer human activities. 
Finally, smart phones are or will be equipped with accelerometers and/or gyroscopes providing data about physical activities of users: \cite{LW14} suggest a complete system of activity recognition based on smartphone accelerometers with potential application to health monitoring.

\emph{Research work related to data grid models:} Dhillon et al.~\cite{DMM03} have proposed an information-theoretic coclustering approach for two discrete random variables: the loss in Mutual Information $MI(Y_1,Y_2) - MI(Y_1^{M} , Y_2^{M})$ is minimized to obtain a locally-optimal grid with a user-defined number of clusters for each dimension. This is limited to two variables and requires to choose the number of clusters per variable.
%
Going beyond 2D matrices, recent significant progress has been done in multi-way tensor analysis~\cite{STF06}. For instance, \cite{MSF+12} suggest a method for mining time-stamped event sequences and effective forecasting of future events.

To the best of our knowledge, our summarization approach is the only one to combine the following advantages: it is parameter-free, scalable and can be applied to mixed-type attributes (categorical, numerical, thus multiple types of time dimensions). Therefore, the same generic method can be used to analyze network graph data, temporal sequence data and mobility data.


\section{Conclusion}
We have suggested a generic methodology for exploratory analysis of CDRs data. Our method is based on a joint distribution estimation technique providing the user with a summary of the data in a parameter-free way. We have also suggested several criteria for exploring and exploiting the summary at various granularities and highlighting its relevant components. We have demonstrated the applicability of the method on graph data, temporal sequence data as well as user mobility data stemming from CDRs data.
%


\bibliographystyle{splncs03}
\bibliography{biblio_general}

\newpage

\appendix

\section{Data grid models in a nutshell}

Data grid models aim at estimating the joint distribution between several variables of mixed-types (categorical as well as numerical). The main principle is to simultaneously partition the values taken by the variables, into groups/clusters of categories for categorical variables and into intervals for numerical variables. The result is a multidimensional ($K$-D) data grid whose cells are defined by a part of each partitioned variable value set. Notice that in all rigor, we are working only with partitions of variable value sets. However, to simplify the discussion we will sometime use a slightly incorrect formulation by mentioning a ``partition of a variable'' and a ``partitioned variable''.

In order to choose the ``best'' data grid model $M^{\ast}$ (given the data) from the model space $\mathcal{M}$, we use a Bayesian Maximum A Posteriori (MAP) approach. We explore the model space while minimizing a Bayesian criterion, called cost. The cost criterion implements a trade-off between the accuracy and the robustness of the model and is defined as follows:

\begin{align}
cost(M) = -\log(\underbrace{p(M \mid D)}_{\textrm{posterior}}) \propto  -\log(\underbrace{p(M)}_{\textrm{prior}} \times \underbrace{p(D \mid M)}_{\textrm{likelihood}})\nonumber
\end{align}

Boull\'e~\cite{Bou11} has shown that we can obtain an exact analytic expression of the cost criterion if we consider a data-dependent hierarchical prior (on the parameters of a data grid model) that is uniform (on a combinatorial point of view) at each stage of the hierarchy. Notice that it does not mean that the prior is uniform, thus in our case, the MAP approach is different from a simple likelihood maximization. The cost criterion is then defined as follows.\\

\begin{definition}
A data grid model $M$ is optimal if minimizing the $cost$ criterion defined as follows:
\label{def:cost}
\begin{align}
cost(M)&= \nonumber \\
& \sum_{k \in \mathbb{K_N}} {\log N}
+ \sum_{k \in \mathbb{K_C}} {\log V_k}
+ \sum_{k \in \mathbb{K_C}} {\log B(V_k, J_k)}\\
& + {\log \binom {N + G - 1} {G - 1}}\\
& + \sum_{k \in \mathbb{K_C}} 
    {\sum_{j_k=1}^{J_k} {\log \binom {N_{j_k}^{(k)} + m_{j_k}^{(k)} - 1} {m_{j_k}^{(k)} - 1}}}\\
& + \log N! 
  - \sum_{j_1 = 1}^{J_1} 
    {\sum_{j_2 = 1}^{J_2} 
      {\ldots
        {\sum_{j_K = 1}^{J_K} 
	        {\log N_{j_1 j_2 \ldots j_K}!}
	        }}}\\
& + \sum_{k \in \mathbb{K}} {\sum_{j_k=1}^{J_k} {\log N_{j_k}^{(k)}!}} 
  - \sum_{k \in \mathbb{K_C}} {\sum_{v_k=1}^{V_k} {\log n_{v_k}^{(k)}!}}
\end{align}
where $N$ is the number of points in the data, $K$ the number of variables, $\mathbb{K}=\{X_1, X_2,\ldots, X_K\}$ the set of variables, $\mathbb{K_N}$ the subset of numerical variables, $\mathbb{K_C}$ the subset of categorical variables, $V_k (k \in \mathbb{K_C})$ the number of values of categorical variable $X_k$, $J_k$ the size of the univariate partition of variable $X_k$, $G = \prod_{k=1}^K {J_k}$ the number of cells of the grid, $m_{j_k}^{(k)} (k \in \mathbb{K_C})$ the number of values of the group/cluster $j_k$ of categorical variable $X_k$, $n_{v_k}^{(k)} (k \in \mathbb{K_C})$ the number of points with value $v_k$ of categorical variable $X_k$, $N_{j_k}^{(k)} (k \in \mathbb{K})$ the number of points in the interval (or value group) $j_k$ of variable $X_k$, $N_{j_1 j_2 \ldots j_K}$ the number of points in the cell $(j_1, j_2, \ldots, j_K)$ of the grid.\\
\end{definition}

The terms of the first three lines stand for the a priori probability of the grid model and constitute the regularization term of the model: complex models (with many clusters for categorical variables and/or many intervals for numerical variables) are penalized. The last two lines stand for the likelihood of data given the parameters of the model: models that are closest to the data are preferred. The extreme case where we have at most one point per cell will maximize the likelihood but we get a very low a priori probability of the grid model, thus a high cost value. The other extreme case, i.e., the null model $M_{\emptyset}$, is when we have only one cell: we have high a priori probability but very low likelihood, thus high cost value. Grids with low cost value indicate a high a posteriori probability $p(M \mid D)$ and are those of interest because they achieve a balanced trade-off between accuracy and generality/simplicity. In terms of information theory, negative logarithm of probabilities can also be interpreted as code length~\cite{Sha48}: here, according to the Minimum Description Length principle (MDL)~\cite{Gru07}, the cost criterion can be interpreted as the code length of the grid model plus the code length of the data given the grid model. Then a low cost value also means a high compression of the data using grid model $M$.\\


\noindent \textbf{Optimization algorithm.} The optimization of data grid is a combinatorial problem: the number of possible partitions of $n$ values of a categorical variable is equal to the Bell number $B(n) = \frac{1}{e}\sum_{k=1}^{\infty}\frac{k^n}{k!}$ and the number of discretizations of $N$ values is $2^N$. Obviously, an exhaustive search is unfeasible and as far as we know, there is no tractable optimal algorithm. Therefore the $cost$ criterion is optimized using a greedy bottom-up strategy whose main principle is described in pseudo-code Algorithm~\ref{algo:gbum}. We start with the finest grained data grid, that is made of the finest possible univariate partitions (for all variables), i.e., based on single value intervals or clusters. Then, we evaluate all merges between clusters and adjacent intervals and perform the best merge if the $cost$ criterion decreases after the merge. We iterate until there is no more improvement of the $cost$ criterion.
%

\begin{algorithm}

  \caption{\textsc{khc}: Data grid optimization}
   \label{algo:gbum}
\SetKwInOut{Input}{Input}
\SetKwInOut{Output}{Output}
\SetFuncSty{textsc}
\SetArgSty{texttt}
\SetDataSty{texttt}
\SetKwComment{Comment}{}{} 

\Input{$M$ Initial data grid solution }
\Output{$M^{\ast}, cost(M^{\ast})\leq cost(M)$ final data grid solution with improved $cost$}
%

$M^{\ast} \leftarrow M$\;\nllabel{algo:gbum:initgrid}

\While{improved data grid solution}{ \nllabel{algo:gbum:stopping}
	$M' \leftarrow M^{\ast}$\; \nllabel{algo:gbum:inittgrid}
	\ForAll{Merge $m$ between two clusters or two intervals}{
		$M^+ \leftarrow M^{\ast} + m$ \Comment*[r]{//consider merge $m$ for grid $M^{\ast}$}
		\If{$cost(M^+) < cost(M')$}{
			$M' \leftarrow M^+$\;
		}
	}
	\If{$cost(M') < cost(M^{\ast})$}{
		$M^{\ast} \leftarrow M'$ \Comment*[r]{// Improved grid solution}
	}
}
\Return{$M^{\ast}$}
\end{algorithm}
In the following, to alleviate the notations, without loss of generality, we consider the 3D case with e.g., two categorical variables with $n$ (respectively $a$ values) and one numerical variable with potentially $N$ values (i.e., the total number of data points).\\
A straightforward implementation of the greedy heuristic remains a hard problem since each evaluation of the $cost$ criterion for a grid $M$ requires $O(naN)$ time, given that the initial finest grid is made of up to $n\times a\times N$ cells. Furthermore, each step of algorithm~\ref{algo:gbum} requires $O(n^2)$ (resp. $O(a^2)$, $O(N)$) evaluations of merges of clusters and intervals; and there are at most $O(n+a+N)$ steps from the finest grained model to the null model. The overall time complexity is bounded by $O(naN(n^2+a^2+N)(n+a+N))$. In~\cite{Bou11}, it has been shown that further optimizations allow to reduce the time complexity to $O(N\sqrt{N}\log N)$. Advanced optimizations combined with sophisticated algorithmic data structures mainly exploits \textit{(i)} the sparseness of the grid, \textit{(ii)} the additivity property of the $cost$ criterion and \textit{(iii)} starts from non-maximal grained grid models using pre and post-optimization heuristics:
\begin{itemize}
\item[\textit{(i)}] In practice data sets represented by 3D points are sparse. Among the $O(naN)$ cells of the grid, at most $N$ cells are non-empty. The contribution of empty cells to the $cost$ criterion in definition~\ref{def:cost} is null, thus each evaluation of a data grid may be performed in $O(N)$ time through advanced algorithmic data structures.
\item[\textit{(ii)}] The additivity of the $cost$ criterion stems from the data-dependent hierarchical prior of criterion. It means that it can be split in a hierarchy of components of the grid model: the variables, then the parts (clusters or intervals) and finally cells. The additivity property allows to evaluate all merges between intervals or clusters in $O(N)$ time. Moreover, the sparseness of the data set ensures that the number of revaluations (after the best merge is performed) is small on average.
\item[\textit{(iii)}] Instead of starting from the finest grained grid, for tractability concern, the algorithm starts from grids with at most $O(\sqrt{N})$ clusters or intervals. Dedicated preprocessing and postprocessing heuristics are employed to locally improve the initial and final solutions produced by algorithm~\ref{algo:gbum}. In these heuristics, the $cost$ criterion is post-optimized alternatively for each variable while the partitions of the others are fixed, by moving values across clusters and moving interval boundaries for the numerical variables. 
\end{itemize}
The optimized version of algorithm~\ref{algo:gbum} is now time-efficient but may lead to a local optimum. To alleviate this concern, we use the Variable Neighborhood Search (VNS) meta-heuristic~\cite{HM01}. The main principle consists of multiple runs of the algorithms using various random initial solutions (we consider 10 rounds of initialization): it allows anytime optimization -- the more you optimize, the better the solution -- while not growing the overall time complexity of algorithm~\ref{algo:gbum}. Full details of the optimization techniques are available in~\cite{Bou11}.


\section{Conclusion}
Data grid models~\cite{Bou11} is an effective method for estimating the joint distribution between several variables of mixed-types and is available under the name of Khiops Coclustering at \url{http://www.khiops.com}. It has also been presented as a demo on several real-world case studies, see e.g.~\cite{GBG+14}.


\end{document}